\documentclass[twocolumn,showpacs]{revtex4}
\usepackage[dvips]{graphicx}

\begin{document}

\title{
In-plane anisotropy on the transport properties 
in the modulated Bi$_2$O$_2$-based conductors Bi-2212 and Bi-Sr-Co-O
}

\author{
T. Fujii$^{ab}$%
\footnote{Corresponding author. Dr. T. Fujii 
Tel.: +81-3-5286-3116; Fax: +81-3-5286-3116. 
{\it E-mail address}: fujii@htsc.phys.waseda.ac.jp}, 
I. Terasaki$^{ab}$, T. Watanabe$^{c}$, and A. Matsuda$^{d}$
}

\address{
$^a$Department of Applied Physics, Waseda University,
Tokyo 169-8555, JAPAN \\
$^b$PRESTO, JST, Kawaguchi 332-0012, JAPAN\\
$^c$NTT Photonics Laboratories, 
$^d$NTT Basic Research Laboratories, \\
3-1 Morinosato Wakamiya, Atsugi-shi, Kanagawa 243-0198, JAPAN
}

\date{Proceedings of ISS 2001}

\begin{abstract}
We investigated the in-plane anisotropy on the resistivity and thermopower of 
the Bi$_2$Sr$_2$CaCu$_2$O$_{8+\delta}$ (Bi-2212) and Bi-Sr-Co-O (BiCo) single 
crystals. In Bi-2212, the $b$-axis resistivity is higher than the $a$-axis 
resistivity, and is expressed as a sum of the $a$-axis resistivity and an 
additional residual resistivity. A downward deviation due to  pseudogap is 
observed below a characteristic temperature $T^*$, which is isotropic in the 
form of conductivity. These results suggest that the modulation structure 
along the $b$-axis works as an anisotropic scattering center, but does not 
affect the pseudogap formation. On the other hand, the anisotropy of the 
resistivity and the thermopower in Pb-doped BiCo is substantial, probably 
owing to the misfit structure between the hexagonal CoO$_2$ layer and the 
rock salt Bi$_2$O$_2$ layer. However, the anisotropy in the resistivity in 
Pb-free BiCo is very small, suggesting that the in-plane anisotropy is 
averaged by the modulation structure, whose direction is tilted by 45$^\circ$ 
from the $a$- and $b$-axes.
\end{abstract}

\keywords{
Bi$_2$Sr$_2$CaCu$_2$O$_{8+\delta}$, cobalt oxide, In-plane 
resistivity, Thremopower, In-plane anisotropy
}

\pacs{74.25.Fy, 74.72.Hs, 74.80.Dm}

\maketitle

\section{Introduction}
Bi$_2$Sr$_2$CaCu$_2$O$_{8+\delta}$ (Bi-2212) and Bi$_2$Sr$_2$Co$_2$O$_x$ 
(BiCo) commonly have a Bi$_2$O$_2$ block layer, in which the lattice is 
modulated by the excess oxygen. Although these two materials have a similar 
crystal structure of alternating stacks of the Bi$_2$O$_2$ layer and the 
CuO$_2$ / CoO$_2$ layer, they show quite different physical properties; 
Bi-2212 is known as a high-temperature superconductor, while BiCo is known as 
a thermoelectric material~\cite{terra1}. High resolution transmission electron 
microscope studies of the Bi-based superconductors (including Bi-2212) have 
clarified that they have strongly modulated structure along $b$-axis and that 
modulation periodicity is reported about 4.8$b$ for Pb-free 
Bi-2212~\cite{matsui1}. This modulated structure disappears in Pb-doped 
Bi-2212~\cite{matsui2}. BiCo was first thought as having a structure 
isomorphic to that of Bi-2212. However, it turned out to have a misfit layer 
structure along the $b$ axis, which consists of alternating stacks of the 
rock-salt Bi$_2$O$_2$ layers and the hexagonal CoO$_2$ layers~\cite{raveau}. 
The modulation structure also exists in the pure BiCo although its direction 
is tilted by 45$^\circ$ from $a$- and $b$-axes in contrast with Bi-2212. Since 
the CuO$_2$ and CoO$_2$ layers are responsible for electric conduction, it is 
interesting how the electric properties are affected by the modulation and the 
misfit structure. Here, we report on the in-plane anisotropy on the resistivity
 and thermopower of the two compounds.

\begin{figure}
 \begin{center}
  \includegraphics*[width=7cm, clip]{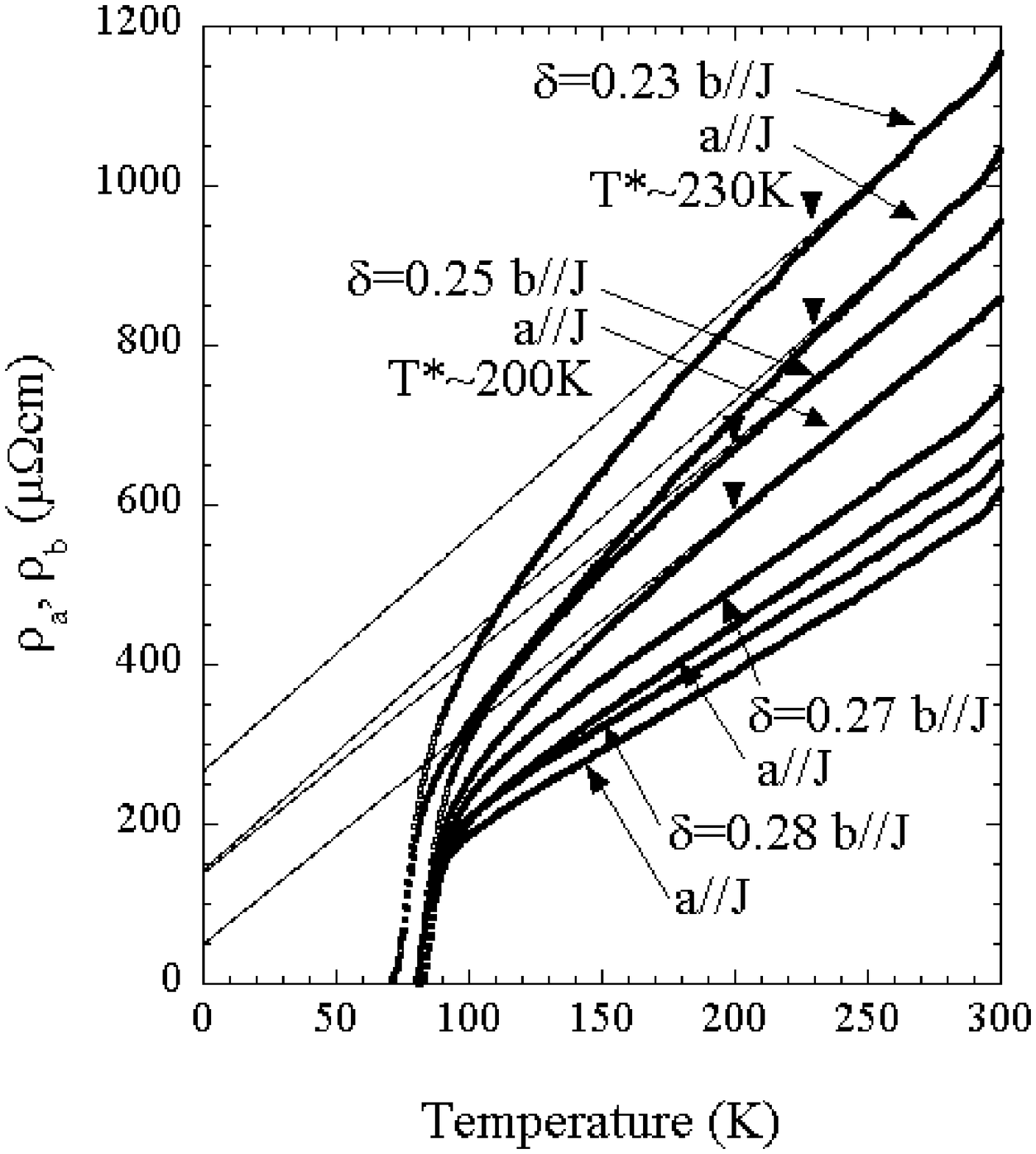} 
 \end{center}
 \caption{
 $a$-axis resistivity $\rho_{a}$ and $b$-axis resistivity $\rho_{b}$ 
 of Bi$_{2.2}$Sr$_{1.8}$CaCu$_2$O$_{8+\delta}$ single crystal 
 for various doping levels measured by the Montgomery method.
 }

 \begin{center}
  \includegraphics*[width=7cm, clip]{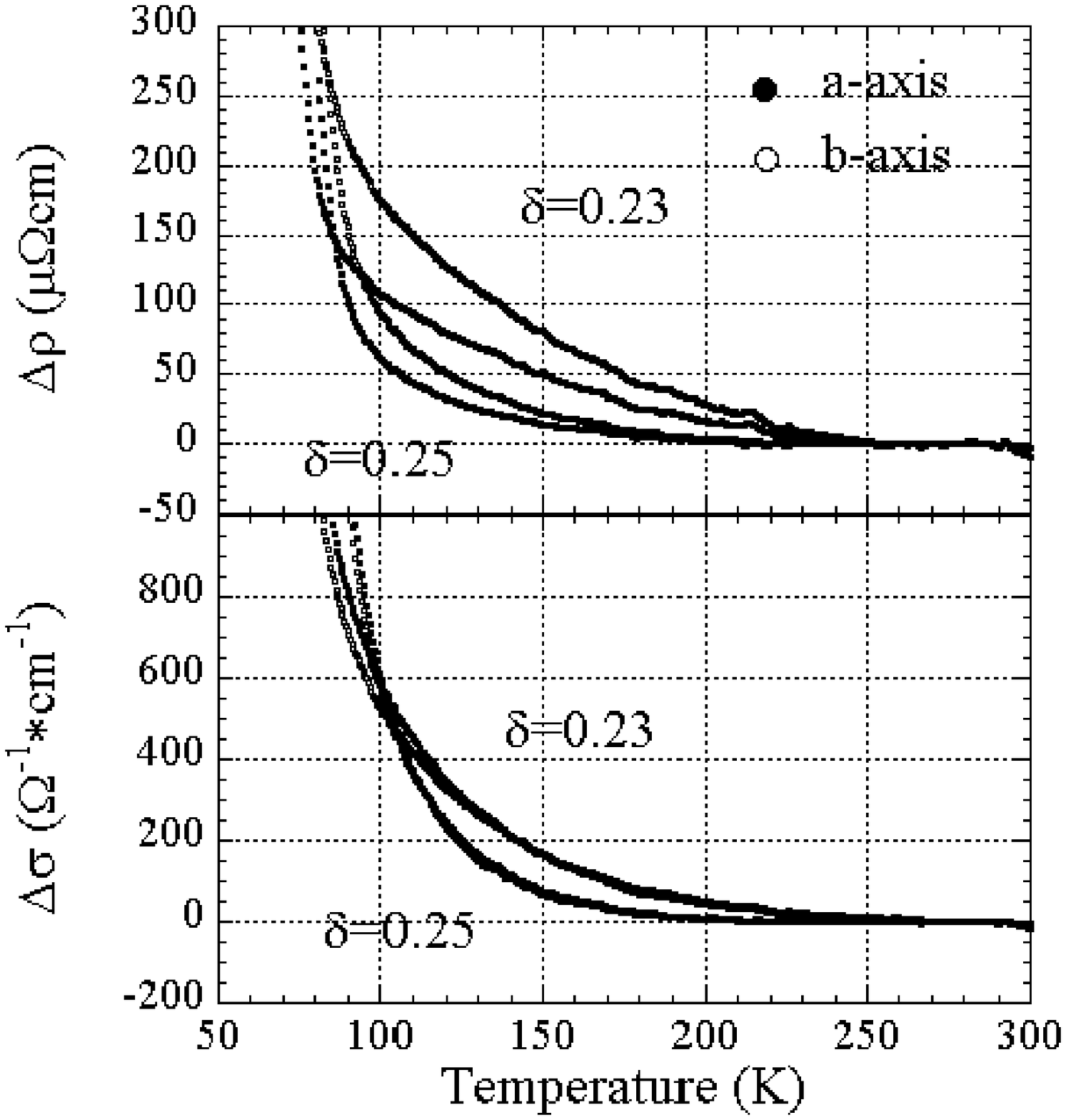} 
 \end{center}
 \caption{
 (a): The deviation from $T$-linear extrapolation ($\Delta\rho_a$ and 
 $\Delta\rho_b$). (b): The change in conductivity 
 $\Delta\sigma$=$\Delta\rho$/$\rho^2$.
 }
\end{figure}

\section{Experimental}
Superconducting compound Bi-2212 with Bi:Sr=2.2:1.8 and thermoelectric material
 Pb-free and 20\% Pb-doped BiCo (Bi$_{1.6}$Pb$_{0.4}$Sr$_2$Co$_2$O$_x$) were 
used in this study. Single crystals of Bi-2212 and BiCo was grown by a 
traveling solvent floating zone method. Starting composition of Bi-2212 and 
BiCo was Bi$_{2.2}$Sr$_{1.8}$CaCu$_2$O$_{8+\delta}$ and 
(Bi,Pb)$_2$Sr$_2$Co$_2$O$_x$ respectively. The growth velocity was adopted to 
be 0.5mm/h, and the growth atmosphere was a mixed gas flow of O$_2$ (20\%) and 
Ar (80\%). Structural analysis for BiCo was performed by using 4-axis X-ray 
diffractmater (MAC Science MXC) and transmission electron microscope (JEOL 
JEM-100CX). Pb-doped BiCo was found to monoclinic symmetry with lattice 
parameter of a=4.91$\AA$, b=5.22$\AA$, c=29.98$\AA$, $\alpha$ and 
$\gamma\approx$90.0$^{\circ}$, $\beta\approx$92.4$^{\circ}$. In TEM 
diffraction patterns, we clearly observed a rock-salt diffraction pattern from 
the BiO layer and a hexagonal diffraction pattern from the CoO$_2$ layer as 
was previously reported in ref. ~\cite{yamamoto}. The Montgomery method was 
adopted to measure the in-plane resistivity of Bi-2212 because they show 
rather small anisotropy between the $a$- and $b$-axes. On the other hand, the 
resistivity of BiCo was measured using a standard four probe method, where the 
current contacts were carefully mounted on the side surface ($ac$- or 
$bc$-surface) so that current density should be homogeneous everywhere in the 
sample. The thermopower was measured using a steady-state technique from 4.2 
to 300K.

\section{Results and discussion}
\subsection{Anisotropy in Bi-2212}
Figure 1 shows the temperature dependence of the $a$- and $b$- axis 
resitstivities ($\rho_a$ and $\rho_b$) for various doping levels. The 
magnitudes and the slopes $d\rho$/$dT$ monotonically decrease with increasing 
oxygen contents $\delta$, indicating that the carriers are actually doped with 
increasing $\delta$~\cite{watanabe}. $\rho_b$ is higher than $\rho_a$, where 
$\rho_{b}$ is roughly expressed as a sum of $\rho_{a}$ and a T-independent 
residual resistivity ($\rho_{0}$) at high temperature. An underdoped sample 
shows a typical downward deviation from high temperature $T$-linear behavior 
below a characteristic temperature $T^*$, which implies pseudogap formation. 
In Fig. 2 (a), we show the magnitudes of the deviation from $T$-linear 
extrapolation ($\Delta\rho_a$ and $\Delta\rho_b$) for oxygen content 
$\delta\sim$0.23, and 0.25. As seen in Fig. 2(a), $\Delta\rho_a$ and 
$\Delta\rho_b$ deviate each other below the pseudogap temperature $T^*$, 
indicating that the decrease of $\rho_b$ is more rapid than that of $\rho_a$. 
Although $\Delta\rho_a$ and $\Delta\rho_b$ are anisotropic in the form of 
resistivity, there is no in-plane anisotropy in the deviation in conductivity 
($\Delta\sigma$=$\Delta\rho$/$\rho^2$) as seen in Fig. 2(b).

\begin{figure}[t]
 \begin{center}
  \includegraphics*[width=7cm, clip]{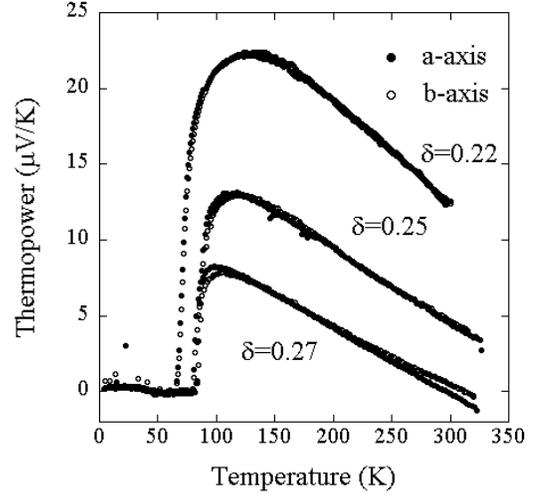} 
 \end{center}
 \caption{
 Thermopower of Bi$_{2.2}$Sr$_{1.8}$CaCu$_2$O$_{8+\delta}$ 
 single crystal for 
 various doping levels measured along the $a$- and $b$- axes.
 }
\end{figure}

Figure 3 shows the temperature dependence of the thermopower in various doping 
levels measured along the $a$- and $b$- axes. The magnitude of the thermopower 
decreases with increasing $\delta$, also indicating that the excess oxygen 
causes the carrier doping. Typical negative slope at high temperature was 
observed in all the doping levels. In contrast with the in-plane resistivities,
 there is no anisotropy in the thermopower within our experimental error of 
10\%. 

These results suggest that the modulation structure along the b-axis works as 
an anisotropic scattering center above $T^*$, but does not affect the 
pseudogap value. In principle, thermopower does not attribute to the 
scattering rate. In YBa$_2$Cu$_3$O$_7$ (YBCO) case, since the CuO chain also 
attributes to the electron conduction, there exists large anisotropy in the 
normal-state resistivity. In superconducting state, angle-resolved 
photoemission spectroscopy (ARPES) measurement reported a significant 
($\sim50\%$) difference in the gap magnitude between $X$ and $Y$ 
point~\cite{shen}. However, this kind of gap anisotropy has not been reported 
in ARPES measurement for Bi-2212. In the superconducting state in the Bi-2212, 
optical conductivity shows an upturn behavior at low frequencies for the 
$b$-axis polarization that is not present in the $a$-axis 
polarization~\cite{MAQ}. This upturn behavior is very simlarly seen in the 
optical conductivity of Zn doped YBCO~\cite{tajima}. This indicates the 
presence of low-lying excitations along the $b$-axis, perhaps due to an 
additional scattering caused by the modulation structure. Thus, an anisotropy 
in the scattering rate exists not only in the normal state but also in the 
superconducting state.

\begin{figure} [t]
 \begin{center}
  \includegraphics*[width=7cm, clip]{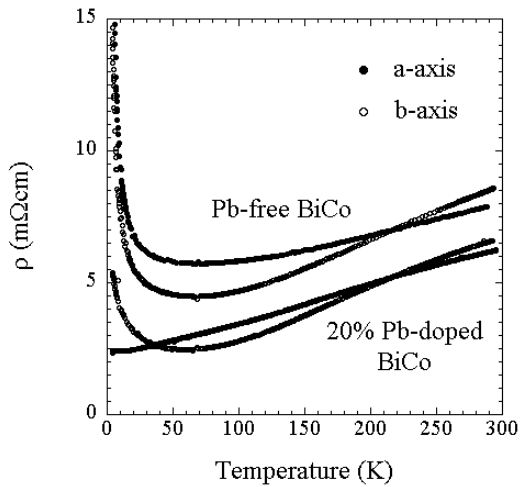} 
 \end{center}
 \caption{
 $a$-axis resistivity $\rho_{a}$ and $b$-axis resistivity $\rho_{b}$ of 
 (Bi,Pb)-Sr-Co-O single crystal measured by the four-probe method.
 }
\end{figure}

\subsection{Anisotropy in BiCo}
Figure 4 shows the temperature dependence of the resistivities for the $a$- 
and $b$-axis directions. Pb-doped BiCo exhibits lower resistivities than 
Pb-free BiCo, suggesting that carrier is doped through the substitution of 
divalent Pb$^{2+}$ for trivalent Bi$^{3+}$. As for the Pb-free sample$, 
\rho_a$ and $\rho_b$ exhibit metallic behavior near room temperature and 
insulating behavior below 50K and the in-plane anisotropy is relatively small. 
On the other hand, the Pb-doped sample shows a significant anisotropy at low 
temperature; $\rho_a$ remains metallic down to 4.2K, whereas $\rho_b$ shows an 
upturn near 50K.

\begin{figure}[b]
 \begin{center}
  \includegraphics*[width=7cm, clip]{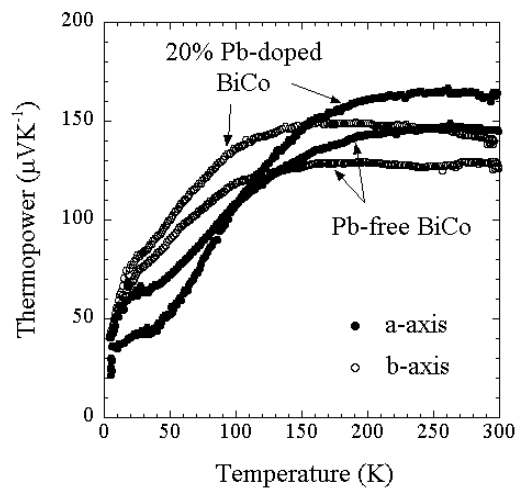} 
 \end{center}
 \caption{
 Thermopower of (Bi,Pb)-Sr-Co-O single crystal measured 
 along $a$- and $b$-axes.
 }
\end{figure}

Figure 5 shows the temperature dependence of the $a$- and $b$-axis thermopower 
for the Pb-free and Pb-doped samples. In contrast with Bi-2212, they exhibit a 
large anisotropy, where the $b$-axis thermopower is about two times larger 
than the $a$-axis thermopower around 50K. At room temperature, the $a$-axis 
thermopower is larger than the $b$-axis thermopower by about 10$\mu$V/K. 
Temperature independent thermopower at high temperature is often explained in 
terms of an entropy per site (the Heikes formula)~\cite{koshibae}, which is in 
principal isotropic. Thus the Heikes formula is unlikely to be applicable to 
the thermopower of BiCo. The large anisotropy in the Pb-doped sample would 
come from the misfit structure along the $b$-axis. On the other hand, in the 
case of the resistivities of the Pb-free sample, anisotropy would be averaged 
by the modulation structure, whose direction is tilted by 45$^\circ$ from the 
$a$- and $b$-axes. It was pointed out that the physical properties of the 
layered cobalt oxides resemble those of the Ce-based compounds~\cite{terra2}. 
The thermopower of CeCu$_2$Ge$_2$ increases with increasing pressure. In the 
Ce-based compound, pressure shrinks the lattice to increase the hybridization 
between the Ce 4$f$ state and the conduction band. The large thermopower of 
BiCo would be also attributed to the increase of hybridization, which comes 
from chemical pressure due to the misfit structure. 

\section{Conclusion}
In conclusion, we investigated the in-plane anisotropy on the resistivity and 
thermopower of the Bi-2212 and Bi-Sr-Co-O. We found in Bi-2212 that the 
modulation structure causes the residual resistivity on $\rho_b$ as 
$\rho_b$=$\rho_a$+$\rho_0$. However, it does not affect the pseudogap 
formation and the thermopower. In the case of Pure BiCo, the in-plane 
anisotropy of the resistivity is averaged by the modulation structure, whose 
direction is tilted by 45$^\circ$ from the $a$- and $b$-axes. On the other 
hand, the misfit structure makes the resistivities and thermopowers of BiCo 
anisotropic. The large anisotropy of BiCo would be attributed to the chemical 
pressure due to the misfit structure.

\end{document}